\documentclass[final,5p,times]{elsarticle}

\usepackage{amssymb}
\usepackage{amsmath}

\usepackage{lineno}

\newcommand{\Rbcs}{R_\mathrm{BCS}}
\newcommand{\Rbcsz}{R_\mathrm{BCS0}}
\newcommand{\Rres}{R_\mathrm{res}}
\newcommand{\Ea}{E_\mathrm{acc}}
\newcommand{\Bp}{B_\mathrm{pk}}

\journal{Nucl Instrum Methods Phys Res A}
\RequirePackage[colorlinks=true,hyperfootnotes=false,bookmarksnumbered=true]{hyperref}
\RequirePackage[all]{hypcap}
\RequirePackage[capitalise,nameinlink]{cleveref}
\RequirePackage{booktabs}
\RequirePackage{ifthen}
\biboptions{sort&compress}

\begin{document}

\begin{frontmatter}



\title{Impact of heat treatments on the performance of low-frequency superconducting quarter-wave resonators at 4.3~K} 
\cortext[cor1]{Corresponding author}

\author[1,2]{Jacob Brown\corref{cor1}} 
\ead{brownjac@frib.msu.edu}
\author[1]{Sang-hoon Kim}
\author[1]{Walter Hartung}
\author[1,2]{Ting Xu}

\affiliation[1]{organization={Facility for Rare Isotope Beams, Michigan State University},
            city={East Lansing},
            state={MI},
            postcode={48824}, 
            country={USA}}
\affiliation[2]{organization = {Department of Physics and Astronomy, Michigan State University},
                city = {East Lansing},
                state = {MI},
                postcode = {48824},
                country = {USA}}

\begin{abstract}
  We applied heat treatments to 80.5~MHz quarter-wave resonators made from bulk niobium and prepared with buffered chemical polishing (BCP\@).  We evaluated their performance at 4.3~K.  We found that a 48~hour, $120^\circ$~C bake-out (``low-temperature bake out'') reduces the surface resistance by a factor of 2 to 3, stemming from a reduction in the Bardeen-Cooper-Schrieffer contribution, consistent with previous findings. This decrease leads to a $38\%$ decrease on average in the medium-field $Q$-slope when compared to cavities which had only BCP\@.  Mechanisms for the change in quality factor with low-temperature baking have been explored.  We observed no improvement in cavity performance after a 3-hour bake-out at $350^\circ$~C (``medium-temperature bake out''), in contrast to observations for higher-frequency cavities.
\end{abstract}

\begin{keyword}
Superconducting RF
\sep SRF Cavity Processing
\sep Ion Linac


\end{keyword}

\end{frontmatter}

%
\ifthenelse{\lengthtest{\columnwidth=\textwidth}}
{\linenumbers}
{\relax}
%
\newlength{\narrowwidth}%
\ifthenelse{\lengthtest{\columnwidth=\textwidth}}
{\setlength{\narrowwidth}{0.8\textwidth}}
{\setlength{\narrowwidth}{\columnwidth}}
%

\section{Introduction}
\label{Introduction}

\begin{figure}
    \centering
    \includegraphics[width=\narrowwidth]{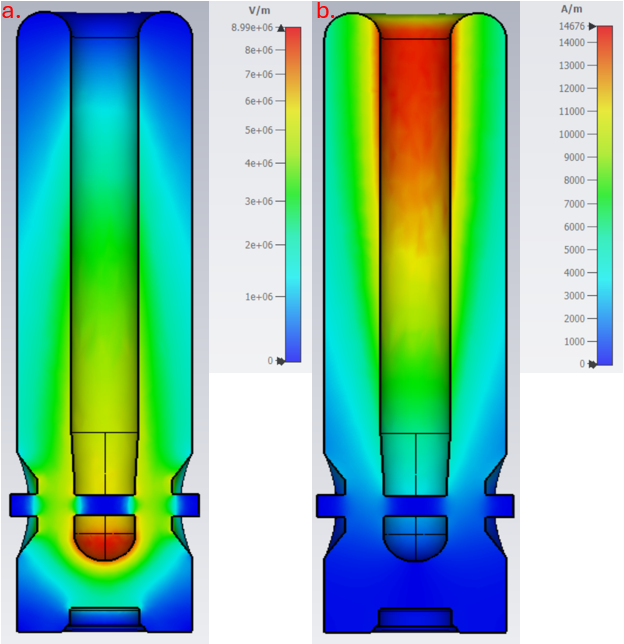}
    \caption{Intensity maps of the magnitude of the electric field (left) and magnetic field (right) for the FRIB $\beta = 0.085$ QWR.  The field amplitudes are normalized to a stored energy of $U = 1$~Joule.}
    \label{fig:qwrs}
\end{figure}

The Facility for Rare Isotope Beams (FRIB) began user operations in May 2022~\cite{mpla37:2230006,osti_og}. The driver linear accelerator (linac) consists of 324 superconducting radio-frequency (SRF) cavities to accelerate light and heavy ions (up to 200 MeV per nucleon in the case of uranium ions).  Four different cavity types are needed, two quarter-wave resonator (QWR) types and two half-wave resonator (HWR) types~\cite{Xu:SRF2017-TUXAA03}.  The second cavity type is a QWR optimized for acceleration of low-velocity beams ($\beta = 0.085$, 80.5 MHz).  The linac includes 92 such cavities in 12 cryomodules.  \Cref{fig:qwrs} shows the field distributions for the accelerating mode calculated with CST Microwave Studio~\cite{ICAP2006:THM2IS03}. 

The quality factor $Q_0$ and surface resistance $R_s$ of an SRF cavity are related via~\cite{padamsee}
\begin{equation}
    Q_0(B_s)=\frac{\omega_0 U}{P_{w}(B_s)}=\frac{G}{R_s(B_s)},
    \label{eq:quality}
\end{equation}
where $U$ is the stored energy, $\omega_0$ is the cavity resonant frequency, $P_w$ is the power dissipation in the cavity, and $G$ is the geometry factor; $R_s$ is the average RF surface resistance of the cavity (weighted according to the distribution of the square of the surface magnetic field $B_s$).  The left equation is generally valid, but the right equation is strictly correct only when there are no additional loss mechanisms in the cavity, such as field emission or multipacting~\cite{graber-FE, Hasan-Padamsee_2001}.  In the absence of additional loss mechanisms, $P_w$ is proportional to $R_s$, so reducing $R_s$ decreases $P_w$, and therefore decreases the load to the cryogenic plant. As such, there has been a sustained effort in the SRF community to reduce $R_s$ and mitigate the $Q$-slope (decrease in $Q_0$ as accelerating gradient $\Ea$ increases).   
Imperfections, foreign particles, and contaminants on the inner surface of the cavity can produce an increase in $R_s$ and may lead to field emission or early thermal breakdown.  Hence careful surface preparation is needed for a high-performance cavity to remove the damage layer from cavity forming and produce the best possible surface.  Standard practice is to remove $\geq 100$~$\mu$m via buffered chemical polishing (BCP\@).  Electro-polishing (EP) is an alternative to BCP which generally produces lower surface roughness~\cite{IEEETNS24:1147to1149, SRF1989:D02, Tian2008TheMO, kelly-bcp, srf-bcp}.

 Baking for $\sim 48$~hours at $120^\circ$~C has been shown to mitigate high-field $Q$-slope at 2~K for electropolished elliptical cavities over a range of frequencies~\cite{Kneisel2000Prelim, gigi-ltb, gigi-hfqs}. This low-temperature baking (LTB) has hence been incorporated as a standard preparation step for the International Linear Collider~\cite{ILC-report} and the European X-ray Free Electron Laser facility~\cite{EuXFEL-report}.

Other treatments have been shown to produce higher $Q_0$ and thus improve the cavity performance. For cavities with frequencies $>1$~GHz, nitrogen doping has been found to be an attractive treatment. Briefly stated, the cavity is heat treated in an ultra-high vacuum (UHV) furnace, nitrogen is introduced into the furnace during the treatment, and then harmful surface nitrides are removed via light EP~\cite{Grassellino_2013, dhakal-ndope-ltb, DHAKAL2020100034}. At GHz frequencies, N-doping results in an increase in $Q_0$ as a function of $\Ea$ (``anti-$Q$-slope''). This anti-$Q$ slope has been shown to have a frequency dependence, and has not been observed in sub-GHz cavities after N-doping, even though $Q_0$ improves~\cite{martinello-field-enhancing, MCGEE-surf}.

Recently, heat treatment at $300$ to $350^\circ$~C for 3 hours (medi\-um-temperature bake, MTB) has been shown to increase $Q_0$ and produce anti-$Q$-slope for 1.3~GHz TeSLA-style cavities at 2~K~\cite{Posen-mtb-tesla, kek-single-mtb, nine-TESLA-mtb, single-TESLA-mtb}. For medium-frequency cavities (500~MHz to 1~GHz), MTB results in higher $Q_0$ at 2~K, in line with N-doping effects for such cavities~\cite{genfa-ln650-mtb, MCGEE-surf}. 

For low-$\beta$ cavities operating at 4.3 K, LTB is often used and is considered helpful to soften multipacting barriers.  For example, LTB was applied for all of the QWRs for SPIRAL2~\cite{longuevergne:tel-00448271, Marchand:SRF2015-WEBA04, SRF2017:THXA08} and was incorporated for ISAC-II at TRIUMF~\cite{Yao:SRF2015-WEA1A03}.  Both groups observed a lessening of the $Q$-slope at $\sim 4.3$~K with LTB.

Recent studies at TRIUMF on multi-mode coaxial resonators prepared with BCP have shown that the effectiveness of MTB and LTB has a frequency dependence~\cite{kolb-multimode, kolb-multimode-bake}.  At 4.3 K, $R_s$ was reduced by MTB for higher-frequency modes ($>$500~MHz), while MTB did not improve the performance for low-frequency modes (220 MHz, 390 MHz). Their work focused on the relationship between the effectiveness of different heat treatments and the frequency of the electromagnetic mode excited in two different coaxial resonators. In our studies, we sought to understand the cause of the improvement in medium-field $Q$-slope for many low-frequency quarter-wave resonators due to LTB as as well as explore the effect of MTB on a FRIB production resonator.

\begin{figure}[htbp]
    \centering
    \includegraphics[width=\columnwidth]{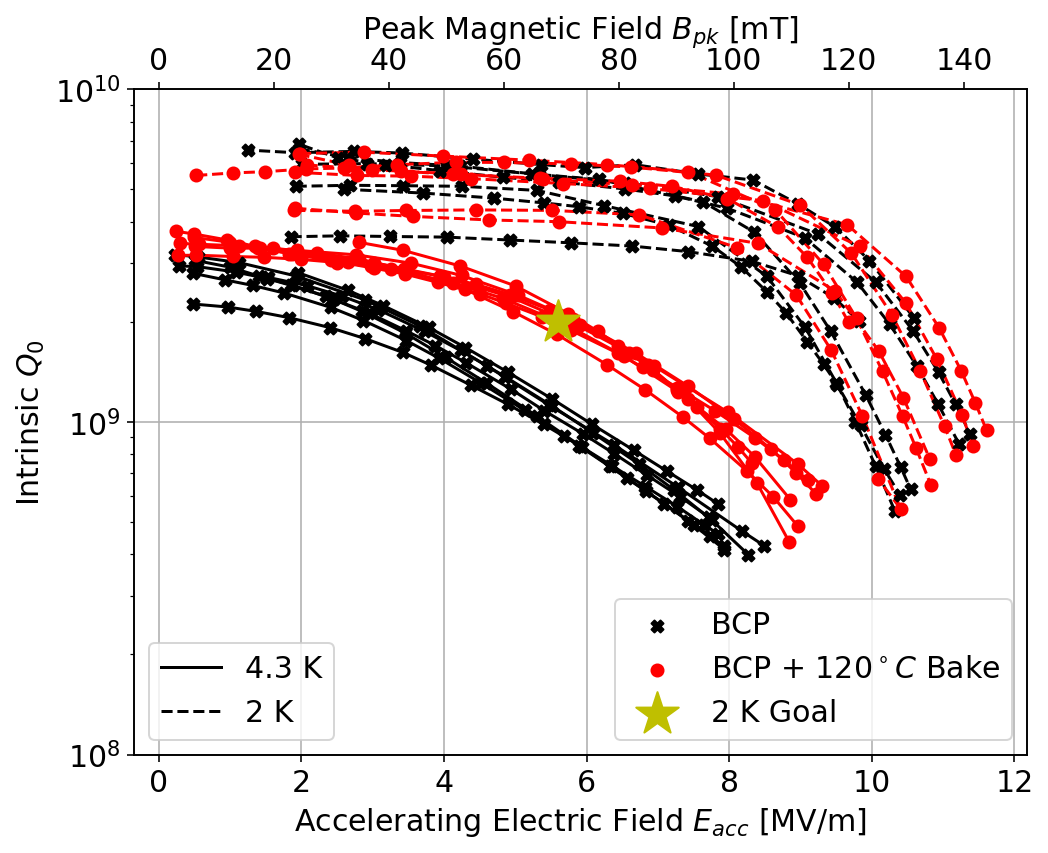}
    \caption{Cavity intrinsic quality factor ($Q_0$) as a function of accelerating gradient ($\Ea$) for several $\beta = 0.085$ QWRs after BCP (black) or BCP+LTB (red) at 4.3~K (solid lines) and 2~K (dashed lines).  The peak surface magnetic field ($\Bp$) is also shown. The star represents the FRIB 2~K cryomodule goal.}
    \label{fig:prelimQ}
\end{figure}

Originally, it was planned to operate the FRIB QWRs at 2~K, co-habitating with superconducting solenoids at 4.3 K in the same cryomodules.  Hence, during FRIB cavity production, LTB was not included as a routine preparation step, as it was not found to be beneficial for 2~K operation~\cite{ZHANG2021165675} (as can be seen in \cref{fig:prelimQ}).  However, during linac commissioning, we found that the QWRs could be stably operated at 4.3~K, which allowed for simpler cryogenic operation of the linac.  (Though the 322~MHz FRIB HWR cryomodules still operate with cavities at 2~K and solenoids at 4.3~K.)  For the present FRIB operating configuration, LTB would provide improved performance for QWRs, so an LTB step is now included for spare and refurbished QWR cryomodules.  As can be seen in \cref{fig:prelimQ}, there is a clear performance improvement for FRIB QWRs at 4.3 K due to LTB.

\section{Cavity Preparation}
FRIB cavities were formed from high-purity niobium sheet by deep drawing and electron beam welding. The residual resistivity ratio (RRR) requirement for the Nb sheet was $>300$. Production cavities were fabricated and jacketed by industrial suppliers and delivered to FRIB for surface preparation and certification testing. Production cavities received a 120~$\mu$m bulk BCP treatment in the FRIB BCP facility.
A $600^\circ$~C hydrogen degassing treatment in an ultra-high vacuum (UHV) furnace at FRIB was done after bulk BCP\@.  Final steps were light BCP and in-clean-room high-pressure water rinsing~\cite{laura-2014}.

  The SRF infrastructure at FRIB~\cite{laura-2012} allows for exploration of different treatments to gauge their effect on cavity performance.  A recent addition is an EP facility~\cite{ethan-2023}, which has been used for FRIB HWRs~\cite{kenji-2023} and prototype elliptical cavities for the proposed FRIB energy upgrade~\cite{kellens}.  The available access ports for the FRIB QWRs make EP treatment impractical for them with the present EP system, thus we did not explore the effect of EP with heat treatments in this work.

\begin{figure}[htb]
    \centering
    \includegraphics[width=\narrowwidth]{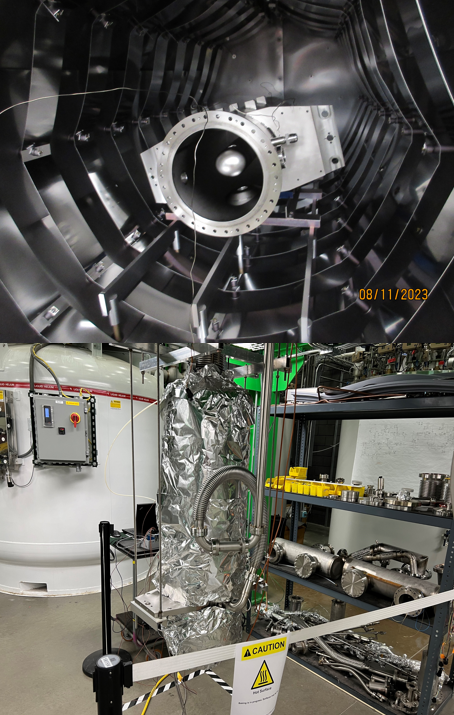}
    \caption{FRIB QWR in the UHV furnace (top) and undergoing in-situ LTB (bottom).}
    \label{fig:bakeSBS}
\end{figure}

\begin{figure}
    \centering
    \includegraphics[width=\columnwidth]{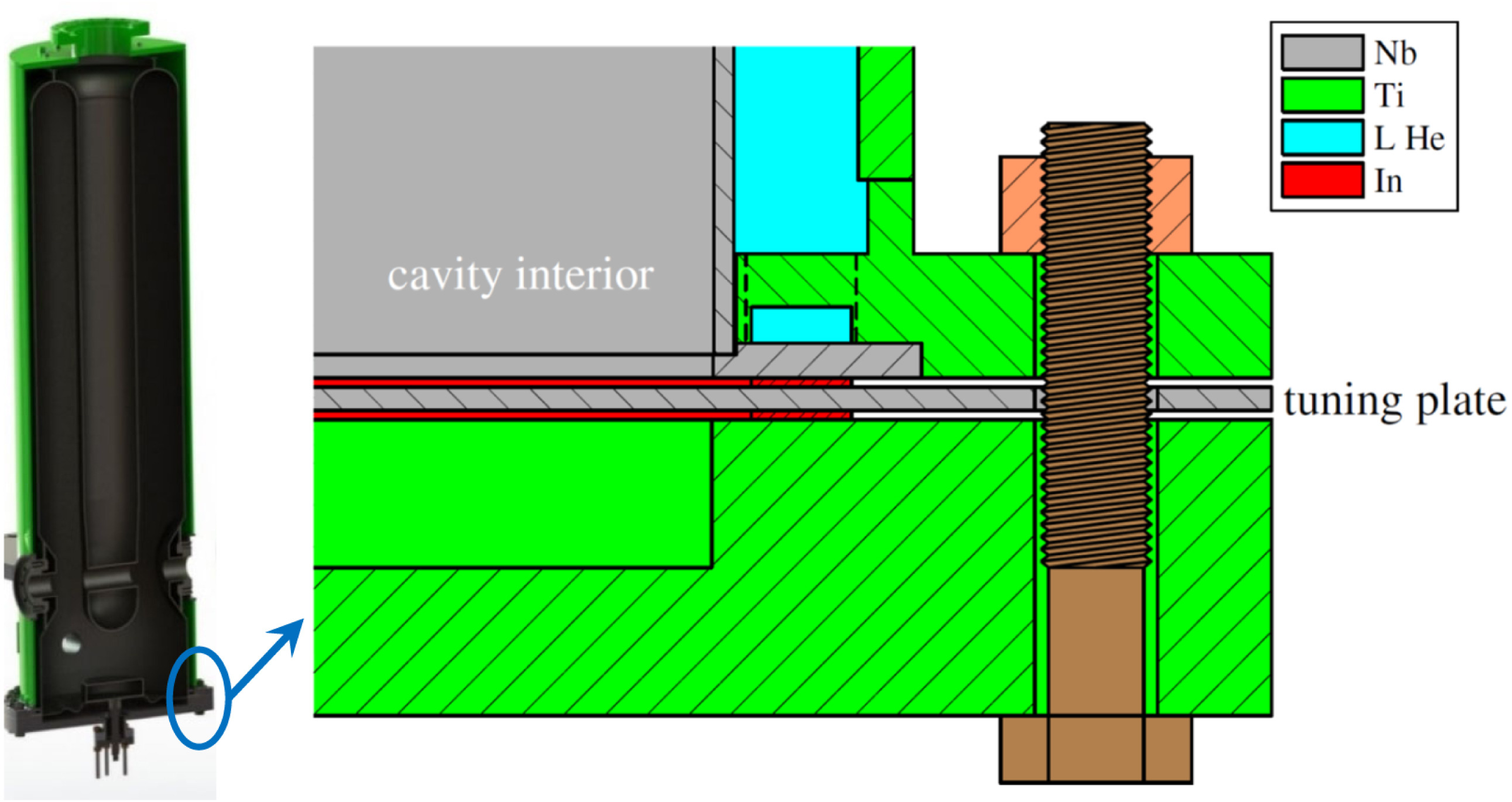}
    \caption{Isometric view of the FRIB $\beta = 0.085$ QWR and schematic of the RF and vacuum seals with a double indium gasket.  Image originally published in Ref.~\cite{ZHANG2021165675}.}
    \label{fig:joint}
\end{figure}

The FRIB UHV furnace can be used for other heat treatments such as LTB and MTB\@. An in-situ bake-out of FRIB cavities after clean-room assembly is also possible using an electric blanket and warm gas flow through the helium vessel.  As FRIB QWRs include an indium gasket, in-situ baking of QWRs cannot be done above about 120$^\circ$~C\@. \Cref{fig:bakeSBS} shows the furnace and in-situ baking setups. A schematic of the QWR RF seal can be seen in \cref{fig:joint}. We note that cavities that undergo LTB in the UHV furnace have the RF surface subsequently exposed to air, while cavities that undergo in-situ LTB do not.

\section{Surface Resistance}
\label{sec3}

The RF surface resistance of a superconductor can be written as the sum of two terms, a term predicted by the Bardeen-Cooper-Schrieffer (BCS) theory of superconductivity, dependent on the temperature $T$ and the RF frequency $f$, and a tem\-per\-a\-ture-independent residual resistance~\cite{padamsee}:
\begin{equation}
    R_s = \Rbcs(T, f) + \Rres\, .
    \label{eq:Rs}
\end{equation}

The BCS term can be expressed to a good approximation by~\cite{padamsee}:
\begin{equation}
    \Rbcs(T, f) = A\left(\frac{f}{\mathrm{1.5~GHz}}\right)^2\frac{T_c}{T}\exp{\left(-\frac{BT_c}{T}\right)}\,,
    \label{eq:RBCS}
\end{equation}
where $T_c$ is the critical temperature of the superconductor (Nb in our case); $A$ and $B$ are coefficients predicted by the BCS theory that are dependent on the London penetration depth $\lambda_L$ and superconducting energy gap $\Delta$. More practically, $A$ and $B$ can be obtained by fitting experimental data. The coefficient $A$ relates to the RRR of the niobium as well. The RRR is correlated with the Nb purity and thermal conductivity which is in turn dependent on the mean-free-path of the electrons in the material; $B$ relates to the superconducting energy gap $\Delta$, which in turn relates to the coherence length of Cooper pairs in the superconductor. A value of $T_c =9.22$~K was used in accordance with findings from Ref.~\cite{gigi-ltb} for niobium after LTB. 

Magnetic flux trapped during the transition from the normal state to the superconducting state can contribute to the residual resistance~\cite{EPAC1992:1295,romanenko-flux-res, gonnella-flux-ipac}. External magnetic field cancellation coils and shielding can be used to try to reduce residual magnetic flux trapping. However, if the cavity and helium jacket are made of different materials, the thermo-electric effect can produce current loops if there are temperature gradients during the during cavity cool-down. The magnetic fields associated with these current loops can be a source of additional trapped flux~\cite{SRF2009:TUPPO053, PRSTAB16:102002}. Other loss mechanisms, including surface impurities, may contribute to $\Rres$ as well. Since it is difficult to describe $\Rres$ in analytical form, our approach is to evaluate $\Rres$ from $Q_0$ at low temperatures where the $\Rbcs$ term in negligible, per \cref{eq:RBCS}, as has been done in previous studies~\cite{romanenko-flux-res, martinello-field-enhancing}.

\section{Methodology}
\begin{figure}
    \centering
    \includegraphics[width=\columnwidth]{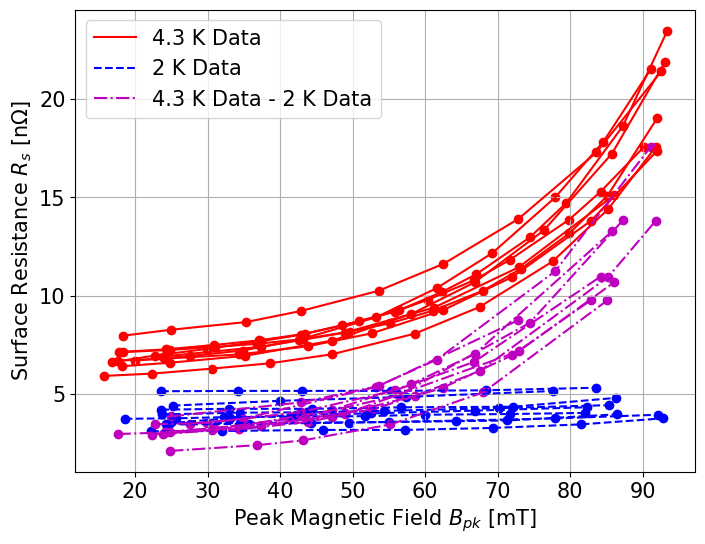}
    \caption{Inferring $\Rbcs$ as a function of field: $R_s$ at 4.3~K (red), $R_s$ at 2~K (blue), and difference (purple).  Results for several FRIB $\beta = 0.085$ QWRs that underwent BCP + LTB are included.}
    \label{fig:bcs-calc}
\end{figure}

We obtain $Q_0$ and $\Ea$ from RF power measurements in continuous-wave mode. We can then calculate $R_s$ from $Q_0$ using \cref{eq:quality} and the $G$ value calculated with CST\@.
Since $f = 80.5$~MHz $\ll 1.5$~GHz for the FRIB QWRs, we expect $\Rbcs\ll \Rres$ at 2~K for our case. This allows us to infer the BCS and residual terms of $R_s$ for different treatments. An example is shown in \cref{fig:bcs-calc}: at 4.3~K (red), the BCS and residual terms both contribute to $R_s$.  At 2~K (blue), the BCS term is negligible, so the 2~K $R_s$ is approximately equal to the residual resistance $\Rres$. Hence the BCS term at 4.3 K can be inferred approximately by subtracting the measured $R_s$ at 2 K from the measured $R_s$ at 4.3 K (purple). 

In \cref{fig:bcs-calc}, we plot the surface resistance as a function of the peak surface magnetic field $\Bp$; $\Bp$ is proportional to $\Ea$, but only represents one location in the cavity. In this sense, using $\Bp$ represents the peak Ohmic losses per unit area on the cavity walls~\cite{padamsee}.

It is clear from \cref{fig:prelimQ} that the biggest difference in $Q_0$ at 4.3~K between unbaked and baked cavities is in the range of $\Ea$ = 2 to 7 MV/m, which corresponds to a $\Bp$ range of about 25 to 87~mT\@. This is within the so-called medium field range of 15 to 95~mT\@. The operating goal for FRIB $\beta=0.085$ QWRs is $\Bp = 69$~mT ($\Ea = 5.6$~MV/m), and thus this medium field range is relevant for FRIB linac operation\@. Hence we will constrain our analysis to this medium field region when comparing differences in cavity preparation.

\section{Measurements and Analysis}
\subsection{Low-field Measurements vs Temperature}

\begin{figure}
    \centering
    \includegraphics[width=\columnwidth]{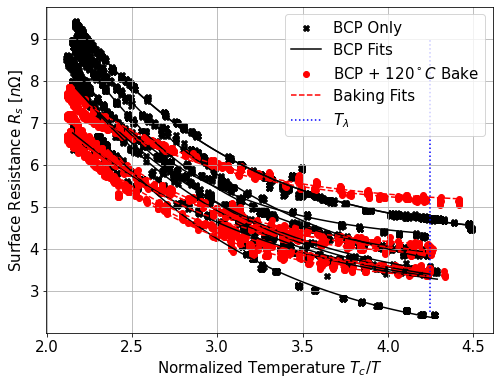}
    \caption{Measured surface resistance at low field as a function of $T_c/T$ for $\beta = 0.085$ QWRs after BCP only (black) or BCP + LTB (red).}
    \label{fig:cdrs}
\end{figure}

Basic information about BCS and residual resistances can be obtained by measuring $R_s$ at constant field as the cavity is cooled from 4.3~K to 2~K. As the temperature decreases, the BCS term decreases until the residual term becomes dominant. An example is shown in \cref{fig:cdrs}. \Cref{eq:Rs} and \Cref{eq:RBCS} can be used to fit the experimental data and obtain $\Rres$, $A$, and $B$. The measurements are taken at a low field, typically $\Ea = 2$~MV/m ($\Bp = 25$~mT) to minimize the field-dependent contribution to $R_s$.

We see from \cref{fig:cdrs} that, at high temperatures such as 4.3~K, $R_s$ decreases after LTB; at low temperature, the $R_s$ values are similar before and after LTB. Thus, at low fields, LTB decreases the BCS term, but not the residual term. 

\begin{figure}[b]
    \centering
    \includegraphics[width=\columnwidth]{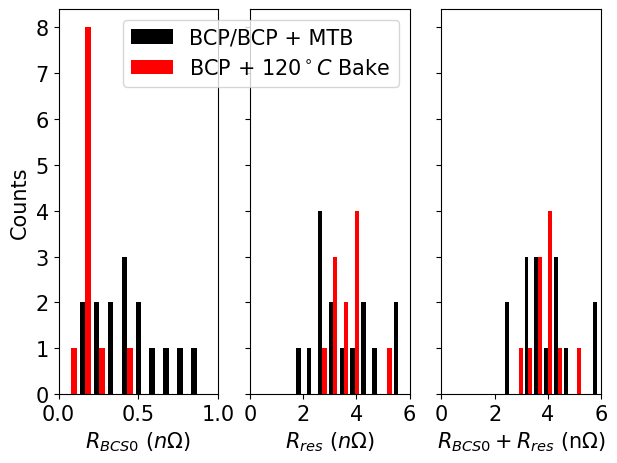}
    \caption{Histograms of $\Rbcsz$ (left), $\Rres$ (middle), and $\Rres+\Rbcsz$ (right) for cavities that underwent BCP/BCP + MTB and those that underwent BCP + 120$^\circ$~C baking.}
    \label{fig:rs0-hist}
\end{figure}

From fitting coefficients $A$, $B$ and the final temperature, we can calculate the low-field value for the BCS resistance at 2~K, $\Rbcsz$. \Cref{fig:rs0-hist} shows the resulting $\Rres$, $\Rbcsz$, and $\Rres+\Rbcsz$ values. It is clear that the LTB cavities (red) tend to have lower $\Rbcsz$ than to BCP'ed cavities (black), with the latter showing more spread.

We note that some extrinsic variables may impact the measured $R_s$ values and may contribute to the scatter seen in \cref{fig:rs0-hist}. It was found that the seating of the indium gasket can affect  $Q_0$~\cite{ZHANG2021165675}. Since this seating does not depend on temperature, imperfect RF contact can contribute to the $\Rres$. Re-torquing of the bottom flange is done before cooling down to mitigate creep in the indium, but there may be some cavity-to-cavity variation due to the joint. 

As mentioned above, another contribution to $\Rres$ is the residual magnetostatic field, which may be trapped when the cavity transitions from normal conducting to superconducting.  Additionally, all of the cold tests are done after helium jacketing and thus thermo-electric currents (discussed in \cref{sec3}) can be generated, leading to flux trapping. There may be variations from one test to another, as we do not carefully control the cool-down rate near $T_c$ for QWR certification tests.

\subsection{Measurements vs Field}

\begin{figure}[htb]
    \centering
    \includegraphics[width=\columnwidth]{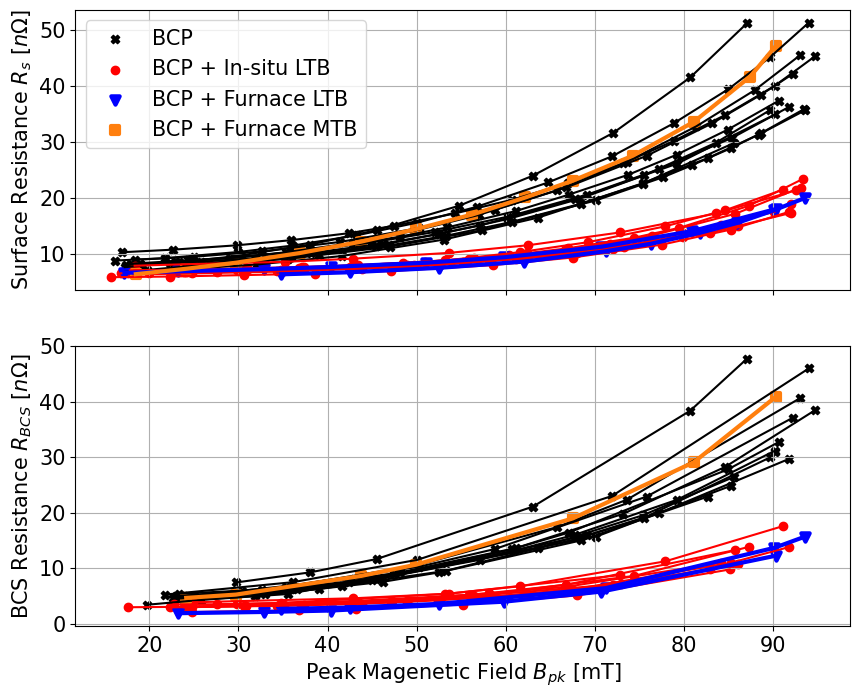}
    \caption{Measured surface resistance $R_s$ (top) and calculated $\Rbcs$ (bottom) at 4.3 K as a function of peak surface magnetic field for several $\beta = 0.085$ QWR tests.}
    \label{fig:rs-comp}
\end{figure}

Surface resistance data from a number of QWR cold tests were collected and surveyed; some results are shown in \cref{fig:rs-comp}. The improvement in high-field performance at 4.3~K with LTB (red circles, blue triangles) relative to BCP alone (black crosses) and BCP + MTB (orange squares) can be seen clearly.
The values of $\Rbcs$ at 4.3~K (\cref{fig:rs-comp}, bottom) are obtained by subtracting the $R_s$ values at 2~K from the $R_s$ values at 4.3~K data, as described above. 

\subsection{Alternative Heat Treatments}

At high field, the BCS resistance is on average 2 to 3 times lower in cavities that underwent BCP + LTB relative to BCP only. Additionally, we see that the BCP only cavities have more spread at high fields than cavities that underwent BCP + LTB. Some additional cases are included in \cref{fig:rs-comp}: we can see that the method of LTB does not affect the outcome, as the 2 furnace LTB cavities (blue triangles) fall soundly alongside the in-situ LTB cavities (red circles). This shows that the performance improvement is not affected by venting after furnace heat treatment. On the other hand, we see that BCP + MTB (orange squares) provides no improvement in $\Rbcs$, performing alongside the standard BCP'ed cavities (black crosses). From the consistency between in-situ LTB and furnace LTB, we can rule out the MTB result being due to a problem with the furnace treatment.

\subsection{Thermal Feedback Model}

When considering possible causes for the reduction in the BCS term with LTB for BCP'ed cavities, one candidate is the thermal feedback (TFB) model~\cite{halbritter-tfb, gigi-ltb, SRF2007:TUP27, padamsee2, triumf_tfbe}. This model considers the impact on the $Q_0$ due to heat transfer through the cavity walls and from the cavity wall to the liquid helium bath. The model predicts
\begin{equation}
    R_s(\Bp)=R_{s0}\left[1+\gamma \left(\frac{\Bp}{B_C}\right)^2\right],
    \label{eq:tfbe}
\end{equation}
where $R_{s0}$ is the low-field value of the surface resistance, $B_C$ is the thermodynamic critical field ($\sim200$ mT), and $\gamma$ is the slope defined for temperature $T_0$ as~\cite{gigi-ltb, SRF2007:TUP27}:
\begin{equation}
    \gamma= \frac{\Rbcsz(T_0)B_C^2\Delta}{2k_BT_0^2}\left[\frac{d}{\kappa}+R_k\right].
    \label{eq:slope}
\end{equation}

In this formulation, $\gamma$ is dependent on the low-field BCS resistance $\Rbcsz$, the thermal conductivity $\kappa$, the Kapitza thermal impedance $R_k$, cavity wall thickness $d$, the superconducting energy gap $\Delta$, Boltzmann's constant $k_B$, and the steady-state helium bath temperature $T_0$. In this formulation, we treat $\gamma$ as a constant and $R_s$ should therefore be a linear function of $(\Bp/B_C)^2$.

\begin{figure}[b]
    \centering
    \includegraphics[width=\columnwidth]{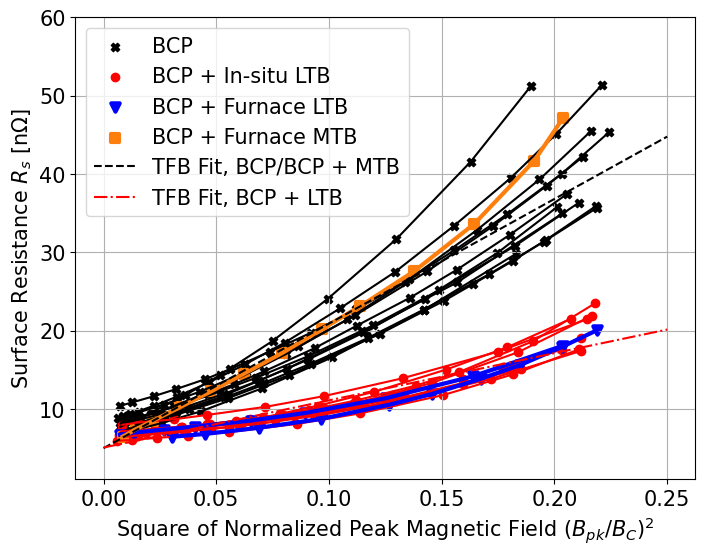}
    \caption{Measured total surface resistance $R_s$ at 4.3~K as a function of $(\Bp/B_C)^2$. Dashed lines: linear fits to the TFB model for the two major groups.}
    \label{fig:rs-bpk2}
\end{figure}

\Cref{fig:rs-bpk2} shows $R_s$ as a function of the square of the normalized peak surface magnetic field for each of the cases.  We can do a linear fit for each cavity and compare the fitted lines to the data. The average of the fitted lines for the BCP and BCP + LTB cases are included in \cref{fig:rs-bpk2} and the fitted parameters are shown in \cref{tab:tfb-fits}.  We note that the measured $R_s$ values tend to increase faster than expected for a linear dependence on $(\Bp/B_C)^2$.
   
\begin{table}
    
    \caption{Fitted parameters for the TFB model: averages for the two major groups of FRIB QWRs.
    \label{tab:tfb-fits}}
    
\begin{center}
    \begin{tabular}{lcr}
        \toprule
         Preparation &  $\gamma$ &  $R_{s0}$ [n$\Omega$]\\
         \midrule
         BCP/BCP+MTB &  $32\pm5$ & $5.0\pm0.7$\\
         BCP+LTB & $12\pm2$ & $5.1\pm0.7$\\
         \bottomrule
    \end{tabular}
\end{center}
    
\end{table}

It is clear that all unbaked and baked cavities have similar low-field surface resistance values ($R_{s0}$). We have already demonstrated that LTB reduces the BCS contribution and increases the residual component. However, the sum of the low field components $R_{s0}=\Rres+\Rbcsz$ is similar for both cases. This is seen in the fitted $R_{s0}$ values (\cref{tab:tfb-fits}) and histogram values of $\Rbcsz + \Rres$ (\cref{fig:rs0-hist}, right).

\section{Discussion}
\subsection{Q-Slope Mitigation Mechanism}

In \cref{tab:tfb-fits}, the slope $\gamma$ decreases by $62\%$ upon low-tem\-per\-a\-ture baking. We now discuss the means by which the slope might be changed by LTB\@.

\subsection{Leading Hypotheses}

In the TFB model, there are two likely suspects: the bulk thermal conductivity and the Kapitza resistance.  Per \cref{eq:slope}, the slope $\gamma$ is proportional to the sum of the thermal impedance of the bulk niobium ($d/\kappa$) and the Kapitza resistance $R_k$. The thermal conductivity $\kappa$ (with the thermal impedance being proportional to its inverse) of Nb can be changed by high-temperature processes such as annealing and degassing~\cite{Fouaidy_2022}, but we do not expect a change due to LTB. 

The Kapitza resistance $R_k$ is the thermal impedance at the liquid helium-niobium interface.  A reduction in $R_k$ would lead to a decrease in  $\gamma$~\cite{gigi-ltb}.
However, if the improvement with LTB was due to a decrease in Kapitza resistance, we would expect no change in cavity performance after repeating light BCP on the inner surface of the cavity, since the outer surface (in contact with the helium bath during the cold test) is not etched.

\subsection{Tests}

\begin{figure}
    \centering
    \includegraphics[width=\columnwidth]{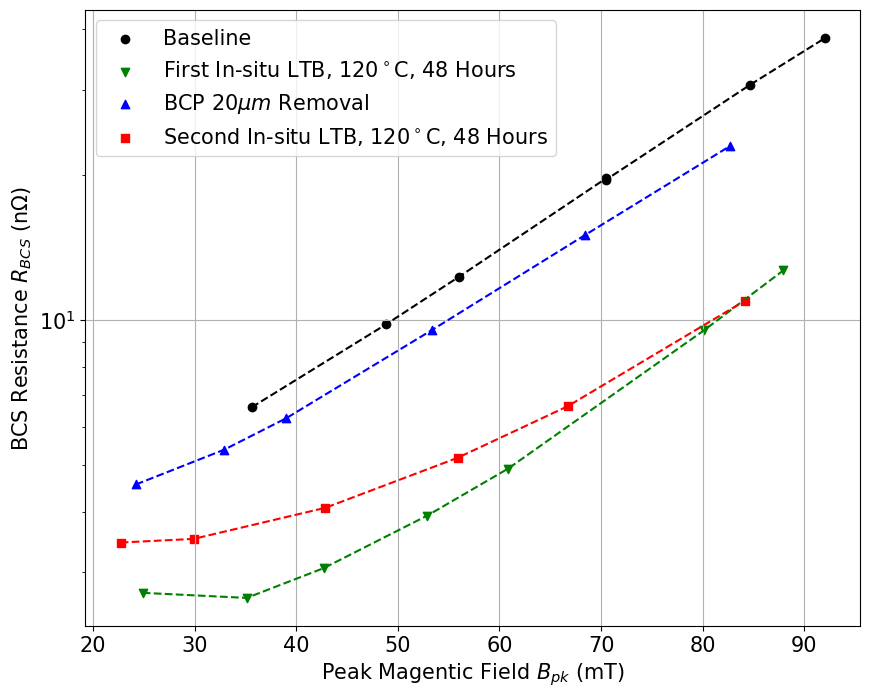}
    \caption{Calculated BCS resistance at 4.3~K as a function of $\Bp$ after different surface preparation steps.  The legend shows the steps preceding each cold test in chronological order.}
    \label{fig:bcp-reset}
\end{figure}

Measurements to test the Kapitza Resistance Change hypothesis on a QWR via different surface preparation steps are shown in  \cref{fig:bcp-reset}.
The performance is improved after each of the two LTB steps, but is worse when light BCP (20~$\mu$m) is applied after the first LTB\@.  The second LTB produces results consistent with the first LTB\@.  This suggests that the improvement in slope from LTB is not associated with a reduction in the Kapitza resistance. 

\subsection{Alternative Hypothesis}

Though we have ruled out 2 likely suspects, the thermal conductivity and Kapitza resistance, we note that the TFB model stipulates that the slope $\gamma$ is directly proportional to the low-field value of the BCS resistance $\Rbcsz$ \cite{halbritter-tfb, gigi-ltb, padamsee2}, as seen in \cref{eq:slope}.  This coefficient comes from the derivative of $R_s$ with respect to $T$.  We saw that LTB reduces $\Rbcsz$, which should lead to a reduction in $\gamma$. Furthermore, this relationship may help to explain the deviation from linearity we see in \cref{fig:rs-bpk2}. Our initial assumption was that $\gamma$ is constant, but as we increase the field amplitude, the cavity wall temperature increases due to the RF wall losses, which leads to an increase in $\Rbcsz$. Thus, we expect an increase in $\gamma$ at high field, which is qualitatively consistent with the measurements.

\subsection{Mean Free Path}

\begin{figure}[b]
    \centering
    \includegraphics[width=\columnwidth]{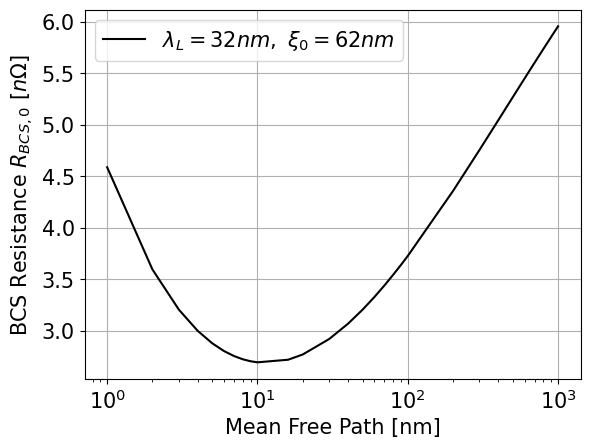}
    \caption{Low-field BCS resistance as a function of the mean free path of normal electrons predicted by SRIMP for an 80.5 MHz cavity at 4.3 K.}
    \label{fig:bcs-mfp}
\end{figure}

Our finding that LTB reduces the zero-field value of $\Rbcs$ may be explained by the Mattis-Bardeen theory and the mean free path. The dependence of the BCS resistance on the mean free path of the normal electrons ($l$) is calculated by the SRIMP code~\cite{halbritter}. 
Some results from SRIMP are shown in \cref{fig:bcs-mfp}. For these calculations, the temperature, London penetration depth $\lambda_L$ and the Cooper-pair coherence length $\xi_0$ are kept constant and only the electron mean free path is varied.
The $\lambda_L$ and $\xi_0$ values from Ciovati, Kneisel, and Myneni~\cite{gigi-ltb} were used.  
We see that, if we increase the mean free path starting with $l=2$~nm, the BCS resistance first decreases and then increases. Baking can allow  diffusion of impurities from the surface into the near-surface region where the RF magnetic field penetrates from the inner surface of the cavity wall. This effect has been studied and documented for higher frequency cavities, but the role of specific contaminants is still under study~\cite{Grassellino_2013, Martinello:2016lrn, Bafia:2021uvd, hu:ipac2024-weps58}. These impurities can reduce the electron mean free path, leading to reduced low-field BCS resistance and thus reduced slope at 4.3~K. 

\section{Conclusion}

We have observed that a low-temperature bake-out (120$^\circ$~C for 48 hours) improves the high-field quality factor of low-frequency FRIB quarter-wave resonators at 4.3 K by lessening the $Q$-slope. In-situ baking after clean-room assembly and furnace low-temperature baking prior to high-pressure water rinsing produce similar results. The increase in $Q_0$ at the design field by about a factor of 2 after LTB is consistent with studies of other low-$\beta$ QWRs such as those of ISAC-II and SPIRAL2.

On the other hand, we found that a medium-temperature bake ($\sim 350^\circ$~C for 3 hours) does not improve the quality factor at 4.3~K\@. This is consistent with low-frequency results from TRIUMF's multi-mode coaxial cavity studies. Cycles of baking and light etching indicate that the improvement from the low-temperature baking is not due to changes in the Kapitza thermal resistance.
The improvement with baking is shown to be a result of reduced BCS resistance, which can be explained as a decrease in the mean free path in the framework of the Mattis-Bardeen theory.

\section{Acknowledgments}
This material is based upon work supported by the U.S Department of Energy, Office of Science (DOE-SC), Office of Nuclear Physics and used resources of the Facility for Rare Isotope Beams Operations, which is a DOE-SC User Facility, under Award Number DE-SC0023633.  Additional support was provided by DOE-SC, Office of High Energy Physics, under Award Number DE-SC0018362.

  \bibliographystyle{elsarticle-num-names} 
  \bibliography{jbrown_ltb_mtb_qwr.bib}

@STRING(jacowpub="{JACoW}")

@STRING(apspub="{APS}")

@article{mpla37:2230006,
author = {Wei, J. and others},
title = {Accelerator commissioning and rare isotope identification at the {Facility for Rare Isotope Beams}},
journal = {Mod. Phys. Lett. A},
volume = {37},
number = {09},
pages = {2230006},
year = {2022},
doi = {10.1142/S0217732322300063}
}

@InProceedings{osti_og,
author = {J. Wei and others},
title = {{FRIB} Transition to User Operations, Power Ramp Up, and Upgrade Perspectives},
doi = {10.18429/JACoW-SRF2023-MOIAA01},
paper = "MOIAA01",
pages = {1--8},
booktitle = "Proc. SRF 2023: 21st Int. Conf. RF Superconductivity, Grand Rapids, MI, USA",
venue = {Grand Rapids, MI, USA},
publisher = jacowpub,
year = {2023}
}

@InProceedings{Xu:SRF2017-TUXAA03,
  author       = {T. Xu and others},
  title        = {{P}rogress of {FRIB} {SRF} Production},
  pages        = {345--352},
  paper        = {TUXAA03},
  doi          = {10.18429/JACoW-SRF2017-TUXAA03},
  booktitle    = {Proc. SRF 2017: 18th Int. Conf. RF Superconductivity, Lanzhou, China},
  venue        = {Lanzhou, China},
  publisher = jacowpub,
  year = {2017},
}

@UNPUBLISHED{ICAP2006:THM2IS03,
AUTHOR="Ulrich Becker",
TITLE="{CST}'s commercial beam-physics codes",
NOTE="presented at ICAP 2006: 9th Int. Computational Accelerator Physics Conference, Chamonix Mont-Blanc, France, Talk THM2IS03",
URL="https://jacow.org/icap06/TALKS/THM2IS03_TALK.PDF",
YEAR=2006
}

@article{ZHANG2021165675,
title = {Certification testing of production superconducting quarter-wave and half-wave resonators for {FRIB}},
journal = {Nucl. Instrum. Methods Phys. Res. A},
volume = {1014},
pages = {165675},
year = {2021},
doi = {10.1016/j.nima.2021.165675},
author = {C. Zhang and W. Hartung and J. Popielarski and K. Saito and S. Kim and W. Chang and T. Xu},
}

@article{kelly-bcp,
author = {M. Kelly and T. Reid},
year = {2017},
month = {01},
pages = {043001},
title = {Surface Processing for Bulk Niobium {SRF} Cavities},
volume = {30},
journal = {Supercond. Sci. Technol.},
doi = {10.1088/1361-6668/aa569a}
}

@inproceedings{srf-bcp,
author = {L. Zhao and C. Reece and M. Kelley},
title = {Genesis of Topography in Buffered Chemical Polishing of Niobium for Application to Superconducting Radio-frequency Accelerator Cavities},
pages = {651--654},
paper = {TUPB108},
doi = {10.18429/JACoW-SRF2017-TUPB108},
booktitle    = {Proc. SRF 2017: 18th Int. Conf. RF Superconductivity, Lanzhou, China},
venue        = {Lanzhou, China},
publisher = jacowpub,
year = {2017}
}

@article{Tian2008TheMO,
  title={The Mechanism of Electropolishing of Niobium in Hydrofluoric-Sulfuric Acid Electrolyte},
  author={Hui Tian and Sean Corcoran and Charles E. Reece and Michael J. Kelley},
  journal={J. Electrochem. Soc.},
  year={2008},
  volume={155},
  pages={D563},
  doi={10.1149/1.2945913}
}

@inproceedings{laura-2012,
    author = {L. Popielarski and C. Compton and L. Dubbs and K. Elliot and A. Facco and L. Harle and I. Malloch and R. Oweiss and J. Ozelis and J. Popielarski and K. Saito},
    title = {Process Developments for Superconducting {RF} Low Beta Resonators for the {ReA3} Linac and {Facility for Rare Isotope Beams}},
    booktitle = {Proc. Linac 2012: 26th Int. Linear Accel. Conf., Tel Aviv, Israel},
    year = {2012},
    paper = {MOPB071},
    pages = {342--344},
    venue = {Tel Aviv, Israel},
    publisher = jacowpub,
    url = {https://jacow.org/LINAC2012/papers/MOPB071.pdf}
}

@inproceedings{laura-2014,
    author = {L. Popielarski and F. Casagrande and C. Compton and T. Elkin and A. Fila and P. Gibson and M. Leitner and I. Malloch and C. Nguyen and R. Oweiss and J. Ozelis and J. Popielarski and C. Thornson and D. Victory and T. Xu},
    title = {{SRF} Highbay Technical Infrastructure for {FRIB} Production at {Michigan} {State} {University}},
    booktitle = {Proc. Linac 2014: 27th Int. Linear Accel. Conf., Geneva, Switzerland},
    year = {2014},
    pages = {954--956},
    paper = {THPP046},
    venue = {Geneva, Switzerland},
    publisher = jacowpub,
    url = {https://jacow.org/LINAC2014/papers/THPP046.pdf}
}

@ARTICLE{IEEETNS24:1147to1149,
author={Shepard, K. W. and Scheibelhut, C. H. and Benaroya, R. and Bollinger, L. M.},
journal={IEEE Trans. Nucl. Sci.},
title={Split Ring Resonator for the {Argonne} Superconducting Heavy Ion Booster},
year={1977},
volume={24},
pages={1147--1149},
doi={10.1109/TNS.1977.4328877}
}

@inproceedings{SRF1989:D02,
title={Preparation and Handling of Superconducting {RF} Cavities},
author={Takaaki Furuya},
year={1989},
booktitle = {Proc. SRF 1989: 4th Workshop RF Supercond., Tsukuba, Japan},
venue = {Tsukuba, Japan},
pages = {305--327},
publisher = jacowpub,
url = {https://www.jacow.org/proceedings/SRF89/papers/srf89d02.pdf}
}

@inproceedings{ethan-2023,
    author = {E. Metzgar and B. Barker and K. Elliot and J. Hulbert and C. Knowles and L. Nguyen and A. Nunham and L. Popielarski and A. Taylor and T. Xu},
    title = {Summary of the {FRIB} Electropolishing Facility Design and Commissioning, Cavity Processing, and Cavity Test Results},
    booktitle = "Proc. SRF 2023: 21st Int. Conf. RF Superconductivity, Grand Rapids, MI, USA",
    venue = {Grand Rapids, MI, USA},
    publisher = jacowpub,
    year = {2023},
    paper = {TUPTB016},
    pages = {419--424},
    doi = {10.18429/JACoW-SRF2023-TUPTB016}
}

@inproceedings{kenji-2023,
    author = {K. Saito and C. Compton and W. Chang and K. Elliot and T. Konomi and S-H. Kim and W. Hartung and E. Metzgar and S. Miller and L. Popielarski and A. Taylor and T. Xu},
    title = {Development of Transformative Cavity Processing---Superiority of Electropolishing on High Gradient Performance Over Buffered Chemical Polishing at Low Frequency (322 {MHz})},
    booktitle = "Proc. SRF 2023: 21st Int. Conf. RF Superconductivity, Grand Rapids, MI, USA",
    venue = {Grand Rapids, MI, USA},
    publisher = jacowpub,
    year = {2023},
    paper = {MOPM026},
    pages = {145--150},
    doi = {10.18429/JACoW-SRF2023-MOPMB026}
}

@article{kellens,
  title = {Medium-velocity superconducting cavity for high accelerating gradient continuous-wave hadron linear accelerators},
  author = {McGee, K. and Kim, S. and Elliott, K. and Ganshyn, A. and Hartung, W. and Metzgar, E. and Ostroumov, P. and Popielarski, L. and Popielarski, J. and Taylor, A. and Xu, T. and Kelly, M. P. and Guilfoyle, B. and Reid, T.},
  journal = {Phys. Rev. Accel. Beams},
  volume = {24},
  pages = {112003},
  year = {2021},
  month = {11},
  publisher = {American Physical Society},
  doi = {10.1103/PhysRevAccelBeams.24.112003}
}

@article{MCGEE-surf,
title = {Advanced surface treatments for medium-velocity superconducting cavities for high-accelerating gradient continuous-wave operation},
journal = {Nucl. Instrum. Methods Phys. Res. A},
volume = {1059},
pages = {168985},
year = {2024},
issn = {0168-9002},
doi = {10.1016/j.nima.2023.168985},
author = {K. McGee and S. Kim and K. Elliott and A. Ganshyn and W. Hartung and P. Ostroumov and A. Taylor and T. Xu and M. Martinello and G.V. Eremeev and A. Netepenko and F. Furuta and O. Melnychuk and M.P. Kelly and B. Guilfoyle and T. Reid}
}

@book{padamsee,
    author = "H. Padamsee and J. Knoblach and T. Hays",
    title = "RF Superconductivity for Accelerators",
    edition = "2nd",
    publisher = "Wiley-VCH Verlag GmbH",
    year = "2008",
    isbn = "978-3-527-40842-9"
}

@article{triumf_tfbe,
  title = {Thermal feedback in coaxial superconducting radio frequency cavities},
  author = {M. McMullin and P. Kolb and Z. Yao and R. Laxdal and T. Junginger},
  journal = {Phys. Rev. Accel. Beams},
  volume = {27},
  pages = {092001},
  year = {2024},
  month = {09},
  publisher = {American Physical Society},
  doi = {10.1103/PhysRevAccelBeams.27.092001}
}

@article{gigi-ltb,
    author = {G. Ciovati},
    title = {Effect of low-temperature baking on the radio-frequency properties of niobium superconducting cavities for particle accelerators},
    journal = {J. Appl. Phys.},
    volume = {96},
    pages = {1591-1600},
    year = {2004},
    month = {08},
    doi = {10.1063/1.1767295},
}

@article{dhakal-ndope-ltb,
  title = {Effect of low temperature baking in nitrogen on the performance of a niobium superconducting radio frequency cavity},
  author = {Dhakal, Pashupati and Chetri, Santosh and Balachandran, Shreyas and Lee, Peter J. and Ciovati, Gianluigi},
  journal = {Phys. Rev. Accel. Beams},
  volume = {21},
  pages = {032001},
  year = {2018},
  month = {03},
  publisher = {American Physical Society},
  doi = {10.1103/PhysRevAccelBeams.21.032001},
}

@article{gigi-hfqs,
  title = {High field {$Q$} slope and the baking effect: Review of recent experimental results and new data on {Nb} heat treatments},
  author = {G. Ciovati and G. Myneni and F. Stevie and P. Maheshwari and D. Griffis},
  journal = {Phys. Rev. ST Accel. Beams},
  volume = {13},
  pages = {022002},
  year = {2010},
  month = {02},
  publisher = {American Physical Society},
  doi = {10.1103/PhysRevSTAB.13.022002}
}

@inproceedings{Kneisel2000Prelim,
  title={Preliminary Experience with ``In-Situ'' Baking of Niobium Cavities},
  author={P. Kneisel},
  year={1999},
  booktitle = {Proc. SRF 1999: 9th Int. Workshop RF Superconductivity, Santa Fe, NM, USA},
  venue = {Santa Fe, NM, USA},
  pages = {328--335},
  paper = {TUP044},
  publisher = jacowpub,
  url = {https://jacow.org/SRF99/papers/TUP044.pdf},
}

@article{kolb-multimode-bake,
  author={P. Kolb and Z. Yao and A. Blackburn and R. Blackburn and D. Hedji and M. McMullin and T.  McMullin and R.E. Laxdal},
  title={Mid-{T} heat treatments on {BCP}’ed coaxial cavities at {TRIUMF}},
  journal={Front. Electron. Mater.},
  volume={3},
  pages={1244126},
  year={2023},
  doi={10.3389/femat.2023.1244126},
}

@article{DHAKAL2020100034,
  title = {Nitrogen doping and infusion in {SRF} cavities: A review},
  journal = {Physics Open},
  volume = {5},
  pages = {100034},
  year = {2020},
  issn = {2666-0326},
  doi = {10.1016/j.physo.2020.100034},
  author = {P. Dhakal}
}

@article{martinello-field-enhancing,
  title = {Field-Enhanced Superconductivity in High-Frequency Niobium Accelerating Cavities},
  author = {Martinello, M. and Checchin, M. and Romanenko, A. and Grassellino, A. and Aderhold, S. and Chandrasekeran, S. K. and Melnychuk, O. and Posen, S. and Sergatskov, D. A.},
  journal = {Phys. Rev. Lett.},
  volume = {121},
  issue = {22},
  pages = {224801},
  numpages = {5},
  year = {2018},
  month = {11},
  publisher = {American Physical Society},
  doi = {10.1103/PhysRevLett.121.224801}
}

@article{single-TESLA-mtb,
author = {Z. Yang and J. Hao and S. Quan and L. Lin and F. Wang and F. Jiao and K. Liu},
title = {Effective medium temperature baking of 1.3 {GHz} single cell {SRF} cavities},
journal = {Physica C},
volume = {599},
pages = {1354092},
year = {2022},
doi = {10.1016/j.physc.2022.1354092}
}

@article{nine-TESLA-mtb,
doi = {10.1088/1361-6668/ac1657},
year = {2021},
month = {08},
publisher = {IOP Publishing},
volume = {34},
number = {9},
pages = {095005},
author = {F. He and W. Pan and P. Sha and J. Zhai and Z. Mi and X. Dai and S. Jin and Z. Zhang and C. Dong and B. Liu and H. Zhao and R. Ge and J. Zhao and Z. Mu and L. Du and L. Sun and L. Zhang and C. Yang and X. Zhang},
title = {Medium-temperature furnace baking of 1.3 {GHz} 9-cell superconducting cavities at {IHEP}},
journal = {Supercond. Sci. Technol.}
}

@inproceedings{kek-single-mtb,
    author = {H. Ito and H. Arakio and K. Takahashio and K. Umemorio},
    title = {Systematic investigation of mid-{T} furnace baking for high-{Q} performance},
    booktitle = {Proc. SRF 2021: 20th Int. Conf. RF Superconductivity},
    pages = {881--884},
    paper = {FROFDV01},
    publisher = jacowpub,
    year = {2021},
    doi = {10.18429/JACoW-SRF2021-FROFDV01}
}

@inproceedings{genfa-ln650-mtb,
    author = {G. Wu and S. K. Chandrasekaran and V. Chouhan and G. Eremeev and F. Furuta and K. McGee and A. Murthy and A. Netepenko and J. Ozelis and H. Park and S. Posen},
    title = {Medium Temperature Furnace Baking of Low-beta 650 {MHz} Five-cell Cavities},
    pages = {158--161},
    paper = {MOPMB030},
    doi = {10.18429/JACoW-SRF2023-MOPMB030},
    booktitle = "Proc. SRF 2023: 21st Int. Conf. RF Superconductivity, Grand Rapids, MI, USA",
    venue = {Grand Rapids, MI, USA},
    publisher = jacowpub,
    year = {2023}
}

@article{Grassellino_2013,
doi = {10.1088/0953-2048/26/10/102001},
year = {2013},
month = aug,
publisher = {IOP Publishing},
volume = {26},
pages = {102001},
author = {A. Grassellino and A. Romanenko and D. Sergatskov and O. Melnychuk and Y. Trenhikam and A. Crawford and A. Rowe and M. Wong and T. Khabiboulline and F. Barkov},
title = {Nitrogen and argon doping of niobium for superconducting radio frequency cavities: a pathway to highly efficient accelerating structures},
journal = {Supercond. Sci. Technol.}
}

@article{Posen-mtb-tesla,
    author = "S. Posen and A. Romanenko and A. Grassellino and O.S. Melnychuk and D.A. Sergatskov",
    title = "Ultralow Surface Resistance via Vacuum Heat Treatment of Superconducting Radio-Frequency Cavities",
    doi = "10.1103/PhysRevApplied.13.014024",
    journal = "Phys. Rev. Appl.",
    volume = "13",
    pages = "014024",
    month = jan,
    year = "2020"
}

@article{kolb-multimode,
  title = {Coaxial multimode cavities for fundamental superconducting rf research in an unprecedented parameter space},
  author = {P. Kolb and Z. Yao and T. Junginger and B. Dury and A. Fothergill and M. Vanderbanck and R.E. Laxdal},
  journal = {Phys. Rev. Accel. Beams},
  volume = {23},
  pages = {122001},
  year = {2020},
  month = {12},
  publisher = {American Physical Society},
  doi = {10.1103/PhysRevAccelBeams.23.122001}
}

@article{romanenko-flux-res,
    author = {A. Romanenko and A. Grassellino and O. Melnychuk and D. A. Sergatskov},
    title = {Dependence of the residual surface resistance of superconducting radio frequency cavities on the cooling dynamics around {$T_c$}},
    journal = {J. Appl. Phys.},
    volume = {115},
    pages = {184903},
    year = {2014},
    month = {05},
    doi = {10.1063/1.4875655}
}

@inproceedings{gonnella-flux-ipac,
    author = {D. Gonnella and J. Kaufman and P.N. Koufalis and M. Liepe},
    title = {{RF} losses from trapped flux in {SRF} cavities},
    booktitle = {Proc. IPAC 2016: 7th Int. Part. Accel. Conf., Busan, Korea},
    year = {2016},
    pages = {2327--2330},
    paper = {WEPMR026},
    publisher = jacowpub,
    venue = {Busan, Korea},
    doi = {10.18429/JACoW-IPAC2016-WEPMR026}
}

@inproceedings{EPAC1992:1295,
author = {C. Vallet and M. Bolor\'{e} and B. Bonin and J. P. Charrier and
B. Daillant and J. Gratadour and F. Koechlin and H. Safa},
title = {Flux Trapping in Superconducting Cavities},
booktitle = {Proc. EPAC 1992: 3rd European Part. Accel. Conf., Berlin, Germany},
pages = {1295--1298},
venue = {Berlin, Germany},
publisher = jacowpub,
year = 1992,
url = {https://www.jacow.org/proceedings/e92/PDF/EPAC1992_1295.PDF}
}

@inproceedings{SRF2009:TUPPO053,
author = {O. Kugeler and A. Neumann and S. Voronenko and W. Anders and
J. Knobloch and M. Schuster and A. Frahm and S. Klauke and
D. Pfl\"{u}ckhahn and S. Rotterdam},
title = {Manipulating the Intrinsic Quality Factor by Thermal Cycling and Magnetic Fields},
booktitle = {Proc. SRF 2009: 14th Int. Conf. RF Supercond., Berlin, Germany},
pages = {352--354},
paper = {TUPPO053},
venue = {Berlin, Germany},
publisher = jacowpub,
year = 2009,
url = {https://jacow.org/SRF2009/papers/TUPPO053.pdf}
}

@article{PRSTAB16:102002,
title = {Impact of cool-down conditions at {${T}_{c}$} on the superconducting rf cavity quality factor},
author = {Vogt, J.-M. and Kugeler, O. and Knobloch, J.},
journal = {Phys. Rev. ST Accel. Beams},
volume = {16},
pages = {102002},
year = {2013},
month = oct,
publisher = apspub,
doi = {10.1103/PhysRevSTAB.16.102002}
}

@techreport{halbritter,
  author       = {Halbritter, J},
  title        = {{FORTRAN}-Program for the Computation of the Surface Impedance of Superconductors},
  institution  = {Institut f\"{u}r Experimentelle Kernphysik, Kernforschungszentrum Karlsruhe},
  address      = {Karlsruhe, West Germany},
  url          = {https://www.osti.gov/biblio/4102575},
  year         = {1970},
  month        = {01}
}

@article{graber-FE,
title = {Reduction of field emission in superconducting cavities with high power pulsed {RF}},
journal = {Nucl. Instrum. Methods Phys. Res. A},
volume = {350},
number = {3},
pages = {572-581},
year = {1994},
issn = {0168-9002},
doi = {10.1016/0168-9002(94)91260-2},
author = {J. Graber and C. Crawford and J. Kirchgessner and H. Padamsee and D. Rubin and P. Schmueser}
}

@article{Hasan-Padamsee_2001,
doi = {10.1088/0953-2048/14/4/202},
year = {2001},
month = apr,
publisher = {},
volume = {14},
number = {4},
pages = {R28},
author = {H. Padamsee},
title = {The science and technology of superconducting cavities for
accelerators},
journal = {Supercond. Sci. Technol.}
}

@inproceedings{SRF2007:TUP27,
    author = {J. Vines and Y. Xie and H. Padamsee},
    title = {Systematic trends for the medium field {Q}-slope},
    booktitle = {Proc. SRF 2007: 13th Int. Workshop RF Superconductivity, Beijing, China},
    year = {2007},
    paper = {TUP27},
    pages = {178--191},
    venue = {Beijing, China},
    publisher = jacowpub,
    url = {https://jacow.org/srf2007/papers/TUP27.pdf}
}

@inproceedings{halbritter-tfb,
    author = {J. Halbritter},
    title = {{RF} residual losses, high electric and magnetic {RF} fields in superconducting cavities},
    booktitle = {Proc. 38th Workshop INFN Eloisatron Project, Erice, Italy},
    month=oct,
    year = {1999},
    pages = {59--79}
}

@book{padamsee2,
    author = "H. Padamsee",
    title = "RF Superconductivity: Science, Technology, and Applications",
    publisher = "Wiley-VCH Verlag GmbH",
    year = "2009",
    isbn = "978-3-527-40572-5"
}

@article{Martinello:2016lrn,
    author = "M. Martinello and A. Grassellino and M. Checchin and A. Romanenko and O. Melnychuck and D.A. Sergatskov and S. Posen and J.F. Zasadzinski",
    title = "Effect of Interstitial Impurities on the Field Dependent Microwave Surface Resistance of Niobium",
    doi = "10.1063/1.4960801",
    journal = "Appl. Phys. Lett.",
    volume = "109",
    pages = "062601",
    year = "2016"
}

@inproceedings{hu:ipac2024-weps58,
    author = {H. Hu and D. Bafia and Y. Kim},
    title = {Decoupling of nitrogen and oxygen impurities in nitrogen doped {SRF} cavities},
    booktitle = {Proc. IPAC 2024: 15th Int. Part. Accel. Conf., Nashville, TN, USA},
    pages = {2825-2828},
    paper = {WEPS58},
    venue = {Nashville, TN, USA},
    publisher = jacowpub,
    month = {05},
    year = {2024},
    issn = {2673-5490},
    isbn = {978-3-95450-247-9},
    doi = {10.18429/JACoW-IPAC2024-WEPS58}
}

@inproceedings{Bafia:2021uvd,
  author       = {D. Bafia and A. Grassellino and A. Romanenko},
  title        = {The Role of Oxygen Concentration in Enabling High Gradients in Niobium {SRF} Cavities},
  booktitle = {Proc. SRF 2021: 20th Int. Conf. RF Superconductivity},
  pages        = {871--875},
  paper          = {THPTEV016},
  publisher    = jacowpub,
  month        = {10},
  year         = {2022},
  issn         = {2673-5504},
  isbn         = {978-3-95450-233-2},
  doi          = {10.18429/JACoW-SRF2021-THPTEV016},
}

@inproceedings{SRF2017:THXA08,
    author = {D. Longuevergne},
    title = {Review of Heat Treatments for Low Beta Cavities: What's So Different from Elliptical Cavities},
    booktitle = {Proc. SRF 2017: 18th Int. Conf. RF Superconductivity, Lanzhou, China},
    pages = {708--714},
    paper = {THXA08},
    venue = {Lanzhou, China},
    publisher = jacowpub,
    doi = {10.18429/JACoW-SRF2017-THXA08},
    year = {2017}
}

@UNPUBLISHED{Yao:SRF2015-WEA1A03,
  author       = {Z. Y. Yao and P. Kolb and R. E. Laxdal},
  title        = {Medium Field {Q}-Slope in Low Beta Resonators},
  venue        = {Whistler, BC, Canada},
  year         = {2015},
  note         = {presented at SRF 2015: Int. Conf. RF Superconductivity, Whistler, BC, Canada, WEA1A03},
  url="https://accelconf.web.cern.ch/SRF2015/talks/wea1a03_talk.pdf"
}

@phdthesis{longuevergne:tel-00448271,
  TITLE = {Etude et test d'un module acc{\'e}l{\'e}rateur supraconducteur pour le projet {SPIRAL2}},
  AUTHOR = {D. Longuevergne},
  URL = {https://theses.hal.science/tel-00448271},
  SCHOOL = {{Universit{\'e} Paris Sud - Paris XI}},
  YEAR = {2009}
  }

@article{Fouaidy_2022,
doi = {10.1088/1757-899X/1241/1/012012},
year = {2022},
publisher = {IOP Publishing},
volume = {1241},
number = {1},
pages = {012012},
author = {M. Fouaidy},
title = {Thermal conductivity of niobium and thermally sprayed copper at cryogenic temperature},
journal = {IOP Conf. Ser.: Mater. Sci. Eng.},
}

@InProceedings{Marchand:SRF2015-WEBA04,
  author       = {C. Marchand and P.-E. Bernaudin and P. Bosland and G. Devanz and R. Ferdinand and Gómez Martínez, Y. and D. Longuevergne and G. Olry and O. Piquet},
  title        = {Performances of {SPIRAL2} Low and High Beta Cryomodules},
  booktitle    = {Proc. SRF 2015: Int. Conf. RF Superconductivity,
                  Whistler, BC, Canada},
  pages        = {967--972},
  paper        = {WEBA04},
  publisher    = jacowpub,
  year         = {2015},
  isbn         = {978-3-95450-178-6},
  url          = {https://accelconf.web.cern.ch/SRF2015/papers/weba04.pdf}
}

@article{ILC-report,
    author = "N. Phinney",
    title = "{ILC reference design report: Accelerator executive summary}",
    reportNumber = "SLAC-PUB-13044",
    journal = "ICFA Beam Dyn. Newslett.",
    volume = "42",
    pages = "7--29",
    year = "2007"
}

@techreport{EuXFEL-report,
    author = "R. Abela and others",
    editor = "M. Altarelli and R. Brinkmann and M. Chergui and W. Decking and B. Dobson and S. Dusterer and G. Grubel and W. Graeff and H. Graafsma and J. Hajdu and J. Marangos and J. Pfluger and H. Redlin and D. Riley and I. Robinson and J. Rossbach and A. Schwarz and K. Tiedtke and T. Tschentscher and I. Vartaniants and H. Wabnitz and H. Weise and R. Wichmann and K. Witte and A. Wolf and M. Wulff and M. Yurkov",
    title = "{XFEL: The European X-Ray Free-Electron Laser. Technical design report}",
    number = "DESY-06-097",
    institution = "DESY",
    doi = "10.3204/DESY_06-097",
    month = "7",
    year = "2006"
}



\end{document}